\documentclass{elsart}
\usepackage{amssymb}
\usepackage{bm}
\usepackage{graphics}
\usepackage{rotating}
\begin{document}
\begin{frontmatter}
\begin{flushright}\normalsize HU-EP-02/14\\
\vskip 1cm
\end{flushright}
\title{
Radiative M1-decays of heavy-light
  mesons in the relativistic quark model} 
\author[hu]{D. Ebert},
\author[hu,scc]{R. N. Faustov},
\author[hu,scc]{V. O. Galkin}
\address[hu]{Institut f\"ur Physik, Humboldt--Universit\"at zu Berlin,
Invalidenstr.110, D-10115 Berlin, Germany}
\address[scc]{Russian Academy of Sciences, Scientific Council for
Cybernetics, Vavilov Street 40, Moscow 117333, Russia}

\begin{abstract}
Radiative magnetic dipole decays of heavy-light vector mesons into
pseudoscalar mesons $V\to P\gamma$ are considered within the
relativistic quark model. The light quark is treated completely
relativistically, while for the heavy quark the $1/m_Q$
expansion is used. It is found that relativistic effects result in
a significant reduction of decay rates. Comparison with previous
predictions and recent experimental data is presented.   
\end{abstract}

\end{frontmatter}

In this letter we consider radiative magnetic dipole (M1) transitions
of the ground state vector ($V$) heavy-light mesons to the
pseudoscalar ($P$) ones, $V\to P\gamma$ (in quark model notations,
$1^3S_1\to 1^1S_0+\gamma$).  For this purpose,
we use the relativistic quark model based on the quasipotential
approach in quantum field theory. Recently, this model has been successfully
applied for the description of different properties of the heavy-light
mesons, such as their mass spectra \cite{egf} and rare radiative
decays \cite{efg}. Our analysis showed that the light quark in the
heavy-light mesons should be treated completely relativistically, while for
the heavy quark it is useful to apply the expansion in powers
of the inverse heavy quark mass $1/m_Q$, which considerably simplifies
calculations. The first and sometimes second order corrections in
$1/m_Q$ are also important for the heavy-light meson decay
description. Analogously, for the radiative decay calculations
considered here,  the
$1/m_Q$ expansion is carried out up to the
second order, and the light quark is treated completely relativistically
(i.~e. without the unjustified expansion in inverse powers of the light
quark mass). It follows from the obtained results that the
relativistic effects give substantial contributions to the
calculated decay rates. 

The radiative $V\to P\gamma$ decay rate is given by \cite{gf}
\begin{equation}
  \label{eq:dr}
  \Gamma=\frac{\omega^3}{3\pi}\left|{\bm{\mathcal  M}}\right|^2,
  \quad {\rm where} \quad \omega=\frac{M_V^2-M_P^2}{2M_V},
\end{equation}
$M_V$ and $M_P$ are the vector and pseudoscalar meson masses. The matrix
element of the magnetic moment $\bm{\mathcal  M}$ is defined by
\begin{equation}
  \label{eq:magm}
  {\bm{\mathcal  M}}=-\frac{i}2\left[\frac{\partial}{\partial{\bf
  \Delta}}\times\left<P\left|{\bf J}(0)\right|V\right>\right]_{\bf\Delta =0},
\qquad {\bf \Delta}={\bf P}-{\bf Q}, 
\end{equation}
where $\left<P\left|J_\mu(0)\right|V\right>$ is the matrix element of the
electromagnetic current between initial vector ($V$) and final
pseudoscalar ($P$) meson states with  momenta ${\bf Q}$ and ${\bf
  P}$ respectively.

We use the relativistic quark model for the calculation of the  matrix
element of the magnetic moment $\bm{\mathcal  M}$ (\ref{eq:magm}). In
our model a meson is described by the wave
function of the bound quark-antiquark state, which satisfies the
quasipotential equation of the Schr\"odinger type in
the center-of-mass frame \cite{egf}:
\begin{equation}
\label{quas}
{\left(\frac{b^2(M)}{2\mu_{R}}-\frac{{\bf
p}^2}{2\mu_{R}}\right)\Psi_{M}({\bf p})} =\int\frac{d^3 q}{(2\pi)^3}
 V({\bf p,q};M)\Psi_{M}({\bf q}),
\end{equation}
where the relativistic reduced mass
is
\begin{equation}
\mu_{R}=\frac{M^4-(m^2_1-m^2_2)^2}{4M^3},
\end{equation}
and 
$b^2(M)$  denotes
the on-mass-shell relative momentum squared
\begin{equation}
{b^2(M) }
=\frac{[M^2-(m_1+m_2)^2][M^2-(m_1-m_2)^2]}{4M^2}.
\end{equation}
Here $m_{1,2}$ and $M$ are quark masses
and a heavy-light meson mass, respectively.

The kernel
$V({\bf p,q};M)$ in Eq.~(\ref{quas}) is the quasipotential operator of
the quark-antiquark interaction. It is constructed with the help of the
off-mass-shell scattering amplitude, projected onto the positive
energy states. An important role in this construction is played
by the Lorentz-structure of the confining quark-antiquark interaction
in the meson.  In
constructing the quasipotential of the quark-antiquark interaction
we have assumed that the effective
interaction is the sum of the usual one-gluon exchange term and the mixture
of vector and scalar linear confining potentials.
The quasipotential is then defined by
\cite{mass}
\begin{equation}
\label{qpot}
V({\bf p,q};M)=\bar{u}_1(p)\bar{u}_2(-p){\mathcal V}({\bf p}, {\bf
q};M)u_1(q)u_2(-q),
\end{equation}
with
$${\mathcal V}({\bf p},{\bf q};M)=\frac{4}{3}\alpha_sD_{ \mu\nu}({\bf
k})\gamma_1^{\mu}\gamma_2^{\nu}
+V^V_{\rm conf}({\bf k})\Gamma_1^{\mu}
\Gamma_{2;\mu}+V^S_{\rm conf}({\bf k}),$$
where $\alpha_s$ is the QCD coupling constant, $D_{\mu\nu}$ is the
gluon propagator in the Coulomb gauge
and ${\bf k=p-q}$; $\gamma_{\mu}$ and $u(p)$ are
the Dirac matrices and spinors.
The effective long-range vector vertex is
given by
\begin{equation}
\Gamma_{\mu}({\bf k})=\gamma_{\mu}+
\frac{i\kappa}{2m}\sigma_{\mu\nu}k^{\nu}, \qquad k^\nu=(0,{\bf k}), 
\end{equation}
where $\kappa$ is the Pauli interaction constant characterizing the
nonperturbative anomalous chromomagnetic moment of quarks. Vector and
scalar confining potentials in the nonrelativistic limit reduce to
\begin{equation}\label{vconf}
V^V_{\rm conf}(r)=(1-\varepsilon)(Ar+B),\qquad
V^S_{\rm conf}(r) =\varepsilon (Ar+B),
\end{equation}
reproducing
\begin{equation}
V_{\rm conf}(r)=V^S_{\rm conf}(r)+
V^V_{\rm conf}(r)=Ar+B,
\end{equation}
where $\varepsilon$ is the mixing coefficient.

The quasipotential for the heavy quarkonia,
expanded in $p^2/m^2$, can be found in Ref.~\cite{mass} and for
heavy-light mesons in \cite{egf}.
All the parameters of
our model, such as quark masses, parameters of the linear confining potential,
mixing coefficient $\varepsilon$ and anomalous
chromomagnetic quark moment $\kappa$, were fixed from the analysis of
heavy quarkonium spectra \cite{mass} and radiative decays \cite{gf}.
The quark masses
$m_b=4.88$ GeV, $m_c=1.55$ GeV, $m_s=0.50$ GeV, $m_{u,d}=0.33$ GeV and
the parameters of the linear potential $A=0.18$ GeV$^2$ and $B=-0.30$ GeV
have the usual quark model values.
In Ref.~\cite{fg} we have considered the expansion of  the matrix
elements of weak heavy quark currents between pseudoscalar and vector
meson ground states up to the second order in inverse powers of
the heavy quark
masses. It has been found that the general structure of the leading,
first,
and second order $1/m_Q$ corrections in our relativistic model is in accord
with the predictions of HQET. The heavy quark symmetry and QCD impose rigid
constraints on the parameters of the long-range potential in our model.
The analysis
of the first order corrections \cite{fg} fixes the value of the
Pauli interaction
constant $\kappa=-1$. The same value of $\kappa$  was found previously
from  the fine splitting of heavy quarkonia ${}^3P_J$- states
\cite{mass}.
The value of the parameter characterizing the mixing of
vector and scalar confining potentials, $\varepsilon=-1$,
was found from the comparison of the second order ($1/m_Q^2$)
corrections in our model \cite{fg} with the same order contributions in HQET.
This value is very close to the one determined from considering radiative
decays of heavy quarkonia \cite{gf}, especially the M1-decays
(e.~g. the calculated decay rate of $J/\Psi\to \eta_c\gamma$ can be
brought in accord with the experiment only with the above value of
$\varepsilon$).

\begin{figure}
  \centering
  \includegraphics{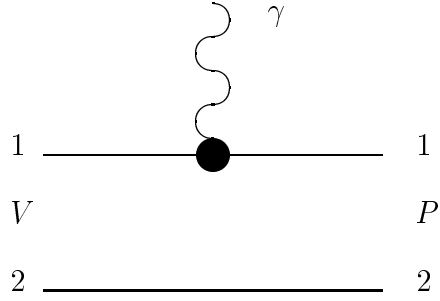}
  \caption{Lowest order vertex function $\Gamma^{(1)}$
corresponding to Eq.~(\ref{gam1}). Radiation only from one quark is shown.}
  \label{fig:1}
\end{figure}
\begin{figure}
  \centering
  \includegraphics{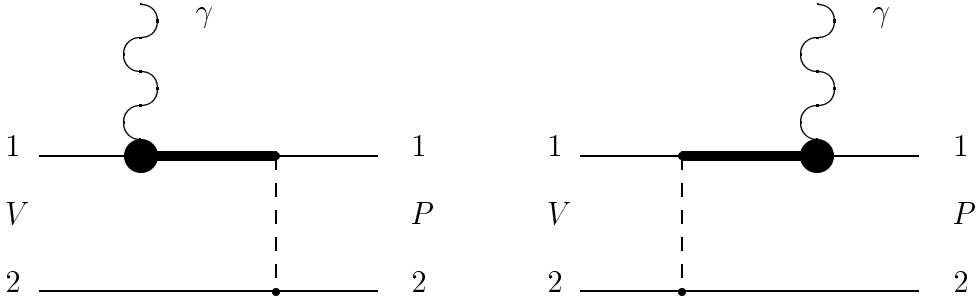}
  \caption{ Vertex function $\Gamma^{(2)}$
corresponding to Eq.~(\ref{gam2}). Dashed lines represent the interaction 
operator ${\mathcal V}$ in Eq.~(\ref{qpot}). Bold lines denote the  
negative-energy part of the quark propagator. As on Fig.~\ref{fig:1},
radiation only from one quark is shown.}
  \label{fig:2}
\end{figure}

In the quasipotential approach,  the matrix element of the electromagnetic
current $J_\mu$ between the states of a vector $V$ meson
and a pseudoscalar $P$ meson has the form \cite{f}
\begin{equation}
\label{mxet}
\langle P \vert J_\mu (0) \vert V\rangle
=\int \frac{d^3p\, d^3q}{(2\pi )^6} \bar \Psi_{P\, {\bf P}}({\bf
p})\Gamma _\mu ({\bf p},{\bf q})\Psi_{V\, {\bf Q}}({\bf q}),
\end{equation}
where $\Gamma _\mu ({\bf p},{\bf
q})$ is the two-particle vertex function and  $\Psi_{V,P}$ are the
meson wave functions projected onto the positive energy states of
quarks and boosted to the moving reference frame.
The contributions to $\Gamma$ come from Figs.~1 and 2.
The contribution $\Gamma^{(2)}$ is the consequence of the
projection onto the positive-energy states. Note that the form
of the relativistic corrections resulting from the vertex function
$\Gamma^{(2)}$ explicitly depends on the Lorentz structure of the
$q\bar q$-interaction.  Thus the vertex function is given by
\begin{equation}
  \label{eq:gam}
  \Gamma_\mu({\bf p},{\bf q})=\Gamma_\mu^{(1)}({\bf p},{\bf q})+
  \Gamma_\mu^{(2)}({\bf p},{\bf q})+ \cdots ,
\end{equation}
where
\begin{equation}\label{gam1}
\Gamma_\mu ^{(1)}({\bf p},{\bf q})=e_1\bar
u_1(p_1)\gamma_\mu  
u_1(q_1)(2\pi)^3\delta({\bf p}_2-{\bf q}_2) +(1\leftrightarrow 2),
\end{equation}
and
\begin{eqnarray}\label{gam2}
\!\!\! \Gamma_\mu^{(2)}({\bf p},{\bf q})&=&e_1\bar u_1(p_1)\bar
u_2(p_2) \biggl\{{\mathcal V}({\bf p}_2-{\bf q}_2)\frac{\Lambda_1^{(-)}(k_1')}{
\epsilon_1(k_1')+ \epsilon_1(q_1)}\gamma_1^0\gamma_{1\mu}
\nonumber\\
& & + \gamma_{1\mu}
\frac{\Lambda_1^{(-)}({k}_1)}{ \epsilon_1(k_1)+\epsilon_1(p_1)}
\gamma_1^0{\mathcal V}({\bf p}_2-{\bf q}_2)
\biggr\}u_1(q_1) u_2(q_2) +(1\leftrightarrow 2). 
\end{eqnarray}
Here $e_{1,2}$ are the quark charges, ${\bf k}_1={\bf p}_1-{\bf\Delta};
\quad {\bf k}_1'={\bf 
q}_1+{\bf\Delta};\quad {\bf\Delta}={\bf P}-{\bf Q}$; 
\[
\Lambda^{(-)}(p)={\epsilon(p)-\bigl( m\gamma ^0+\gamma^0({\bf
\bm{\gamma} p})\bigr) \over 2\epsilon (p)}, \qquad \epsilon(p)=
\sqrt{p^2+m^2}.
\]
It is important to note that the wave functions entering the current
matrix element (\ref{mxet}) cannot  be both in the rest frame.
In the initial $V$ meson rest frame, the final $P$ meson is moving with the recoil
momentum ${\bf \Delta}$. The wave function
of the moving $P$ meson $\Psi_{P\,{\bf\Delta}}$ is connected
with the wave function in the rest frame
$\Psi_{P\,{\bf 0}}\equiv \Psi_{P}$ by the transformation \cite{f}
\begin{equation}
\label{wig}
\Psi_{P\,{\bf\Delta}}({\bf p})
=D_1^{1/2}(R_{L_{\bf\Delta}}^W)D_2^{1/2}(R_{L_{\bf\Delta}}^W)
\Psi_{P\,{\bf 0}}({\bf p}),
\end{equation}
where $R^W$ is the Wigner rotation, $L_{\bf\Delta}$ is the Lorentz boost
from the rest frame to a moving one, and $D^{1/2}(R)$ is
the rotation matrix  in the spinor representation. 

We substitute the vertex functions $\Gamma^{(1)}$  and $\Gamma^{(2)}$
given by Eqs.~(\ref{gam1}) and (\ref{gam2})
in the decay matrix element (\ref{mxet})  and take into account the wave
function transformation (\ref{wig}). To simplify calculations we 
note that the mass of the heavy-light mesons $M_{V,P}$ is large (due
to presence of the heavy quark $M_{V,P}\sim m_Q$)
and carry out the expansion in inverse powers of this mass up to the
second order. Then we calculate the matrix element of the magnetic
moment operator (\ref{eq:magm}) and get\\
(a) for the vector potential
\begin{eqnarray}
  \label{eq:muv}   
\!\!\!\!\!\!\!\!
  {\bm{\mathcal  M}}_V& =& \int \frac{d^3 p}{(2\pi)^3}\bar \Psi_P({\bf
  p}) \frac{e_1}{2\epsilon_1(p)}\Biggl\{\bm{\sigma}_1+
  \frac{(1-\varepsilon)(1+2\kappa)[{\bf p}\times[\bm{\sigma}_1\times
  {\bf p}]]}{2\epsilon_1(p) 
  [\epsilon_1(p) +m_1]} \cr\cr
&&
+\frac{(1-\varepsilon)(1+\kappa)[{\bf p}\times[\bm{\sigma}_2\times
  {\bf p}]]}{\epsilon_1(p) 
  [\epsilon_2(p) +m_2]}\cr\cr
&&-\frac{\epsilon_2(p)}{M_V}\left(1+(1-\varepsilon)
  \frac{M_V-\epsilon_1(p)-\epsilon_2(p)}{\epsilon_1(p)}\right)i\left[{\bf
  p}\times \frac{\partial}{\partial{\bf p}}\right] \cr\cr
&& +\frac1{2M_V}\left[{\bf p}\times\left[{\bf p}\times\left(\frac{\bm{\sigma}_1}
  {\epsilon_1(p)+m_1}
  -\frac{\bm{\sigma}_2}{\epsilon_2(p)+m_2}\right)\right]\right]
\Biggr\}\Psi_V({\bf p}) +(1\leftrightarrow 2),\nonumber\\ 
\end{eqnarray}
(b) for the scalar potential
\begin{eqnarray}
  \label{eq:mus}   
\!\!\!\!\!\!\!\!  
{\bm{\mathcal  M}}_S& =& \int \frac{d^3 p}{(2\pi)^3}\bar \Psi_P({\bf
  p}) \frac{e_1}{2\epsilon_1(p)}\Biggl\{\left(1+
\varepsilon\;
\frac{\epsilon_1(p)+\epsilon_2(p)-M_V}{\epsilon_1(p)}\right)
\cr\cr
&&\times  \left(\bm{\sigma}_1- \frac{\epsilon_2(p)}{M_V}i\left[{\bf
  p}\times\frac{\partial}{\partial{\bf p}}\right] \right)
-  \frac{\varepsilon\;[{\bf p}\times[\bm{\sigma}_1\times {\bf p}]]}{2\epsilon_1(p)
  [\epsilon_1(p) +m_1]} \cr\cr
&& +\frac1{2M_V}\left[{\bf p}\times\left[{\bf p}\times\left(\frac{\bm{\sigma}_1}
  {\epsilon_1(p)+m_1}
  -\frac{\bm{\sigma}_2}{\epsilon_2(p)+m_2}\right)\right]\right]
\Biggr\}\Psi_V({\bf p}) +(1\leftrightarrow 2).\nonumber\\
\end{eqnarray}
Note that the last terms in Eqs.~(\ref{eq:muv}), (\ref{eq:mus}) result
from the wave function transformation (\ref{wig}) from the moving
reference frame to the rest one. It is easy to see that in the
limit $p/m\to 0$ the usual nonrelativistic expression for the magnetic
moment follows.   

Since we are interested in radiative transitions of the 
vector mesons to the pseudoscalar mesons it is possible to evaluate
spin matrix elements
using the relation $\left<\bm{\sigma}_1\right>=-
\left<\bm{\sigma}_2\right>$. Then assuming one quark to be
light $q$ and the other one $Q$ to be heavy and further expanding
Eqs.~(\ref{eq:muv}), (\ref{eq:mus}) in the inverse powers of the heavy
quark mass $m_Q$ up to the second order corrections to the leading
contribution we get\\
(a) for the purely vector potential ($\varepsilon=0$) 
\begin{eqnarray}
  \label{eq:muvex}   
{{\mathcal  M}}_V& =&\frac{e_q}{2m_q}\Biggl\{\left<\frac{m_q}
  {\epsilon_q(p)}\right>-\frac{m_q}3\left<\frac{{\bf
  p}^2}{\epsilon_q(p)[\epsilon_q(p)+m_q]}\left(\frac1{\epsilon_q(p)}+
  \frac1{M_V}\right)\right>\cr\cr
&& +\frac{(1+\kappa)m_q}{3}\left<\frac{{\bf p}^2}{\epsilon_q^2(p)}
  \left(\frac2{\epsilon_q(p)+m_q}-\frac1{m_Q}\right)\right> 
-\frac{m_q}{6M_Vm_Q}\left<\frac{{\bf p}^2}
  {\epsilon_q(p)} \right>\Biggr\}\cr\cr
&&- \frac{e_Q}{2m_Q}\Biggl\{1-\frac{2\left<{\bf
  p}^2\right>}{3m_Q^2}+\frac{1+\kappa}3 \left<\frac{{\bf p}^2}{m_Q}\left(
  \frac1{m_Q} -\frac2{\epsilon_q(p)+m_q}\right)\right>\cr\cr
&&-\left<\frac{{\bf p}^2}{6M_V}\left(\frac1{m_Q}+\frac2{\epsilon_q(p)
  +m_q} \right)\right>\Biggr\},
\end{eqnarray}
(b) for the purely scalar potential ($\varepsilon=1$)
\begin{eqnarray}
  \label{eq:musex}   
{{\mathcal  M}}_S& =&\frac{e_q}{2m_q}\Biggl\{2\left<\frac{m_q}
  {\epsilon_q(p)}\right>-\left<\frac{m_q(M_V-m_Q)}{\epsilon_q^2(p)}
\right>-\frac{m_q}{6M_Vm_Q}\left<\frac{{\bf p}^2}
  {\epsilon_q(p)} \right>\cr\cr
&& +\frac{m_q}{2}\left<\frac{{\bf p}^2}{\epsilon_q^2(p)}
  \left(\frac1{m_Q}-\frac2{3[\epsilon_q(p)+m_q]}- \frac{2\epsilon_q(p)}
  {3M_V[\epsilon_q(p)+m_q]} \right)\right>\Biggr\} 
\cr\cr
&&- \frac{e_Q}{2m_Q}\Biggl\{2-\frac{M_V-\left<\epsilon_q(p)\right>}{m_Q}-
  \frac{\left<{\bf p}^2\right>}{6m_Q}\left(
  \frac1{m_Q} +\frac1{M_V}\right)\cr\cr
&&-\frac1{3M_V}\left<\frac{{\bf p}^2}{\epsilon_q(p)
  +m_q}\right>\Biggr\}.
\end{eqnarray}
Here $\left<\cdots\right>$ denotes the matrix element between radial
meson wave functions. For these matrix element calculations we use the
wave functions of heavy-light mesons obtained in Ref.~\cite{egf}. It
is important to note that
in this reference while calculating the heavy-light meson mass spectra
only the heavy quark was treated using the $1/m_Q$
expansion but the light quark was treated completely
relativistically. 

The values of decay rates of mesons with open flavour calculated on
the basis of  Eqs.~(\ref{eq:dr}),
(\ref{eq:muvex}), (\ref {eq:musex}) are displayed in Table~\ref{tab:1}. In
the second column ($\Gamma^{\rm NR}$) we give predictions for
decay rates obtained in
the nonrelativistic approximation ($p/m\to 0$) for both heavy and
light quarks. In the third ($\Gamma^V$) and fourth ($\Gamma^S$) columns
we  show the results
obtained for the purely vector and scalar confining potentials,
respectively. And in the last column ($\Gamma$) we present predictions
for the mixture of vector and scalar confining potentials
(\ref{vconf}) with the mixing parameter $\varepsilon=-1$. As seen from
this Table relativistic effects significantly influence the
predictions. Their inclusion results in a significant reduction of
decay rates ($\Gamma^{\rm NR}/\Gamma = 2\div 4.5$). Both relativistic
corrections to the heavy quark and the relativistic treatment of the
light quark play an important role. The dominant decay modes of $D^*$
mesons are the strong decay $D^*\to D\pi$, which is considerably
suppressed by the phase space, and the electromagnetic decay $D^*\to
D\gamma$. The corresponding branching ratios are known already for a
long time and listed in PDG tables \cite{pdg}. However, the total
decay rates of $D^*$ mesons were not measured until recently. In
Ref.~\cite{cleoG} CLEO collaboration reported the first measurement of
the $D^{*+}$ decay width $\Gamma(D^{*+})=96\pm 4\pm
22$~keV. Combining this value with the measured $BR(D^{*+}\to
D^+\gamma)=(1.6 \pm 0.4)\%$ \cite{pdg}, the following experimental
value of the decay rate can be obtained: $\Gamma(D^{*+}\to D^+\gamma)=
(1.5\pm 0.6)$~keV. Our model prediction is in agreement with this
experimental value. However, the experimental errors are still large
in order to discriminate the relativistic and nonrelativistic results.
In the case of $B$ mesons the pion emission is kinematically
forbidden, so the dominant
decay mode is electromagnetic decay $B^*\to B\gamma$. None of $B^*$
widths has been measured yet. We also present our predictions for the
radiative decay rate of $B_c^*$ meson, which consists of two heavy
quarks ($b$ and $c$). Therefore the expressions (\ref{eq:muvex}) and
(\ref{eq:musex}) can be further expanded in inverse powers of both
quark masses up to the second order.

\begin{table}
 \caption{Radiative decay rates of mesons with an open flavour (in keV).}
    \label{tab:1}
\centering
    \begin{tabular}{ccccc}
\hline
Decay& $\Gamma^{\rm NR}$ & $\Gamma^V$ & $\Gamma^S$&$\Gamma$\\
\hline
$D^{* \pm}\to D^{\pm}\gamma$& 2.08& 0.60& 0.28 &1.04\\
$D^{* 0}\to D^{0}\gamma$& 37.0& 14.3& 17.4 &11.5\\
$D_s^*\to D_s\gamma$& 0.36&0.13&0.08& 0.19\\
$B^{* \pm}\to B^{\pm}\gamma$& 0.89&0.24&0.29&0.19\\
$B^{* 0}\to B^0\gamma$& 0.27&0.087& 0.101& 0.070 \\
$B_s^*\to B_s\gamma$& 0.132&0.064&0.074&0.054\\
$B_c^*\to B_c\gamma$& 0.073&0.048&0.066&0.033\\
\hline
    \end{tabular}
   
\end{table}

In Table~\ref{tab:2} we compare our predictions for radiative decay
rates of vector heavy-light mesons with other theoretical results. We
show the predictions obtained in quark models
\cite{iv,oh,gr}, in the framework of heavy quark effective theory (HQET)
combined with vector meson dominance (VMD) hypothesis \cite{cfn} and in QCD
sum rules \cite{dn,adip,zyh}. These predictions vary quite
significantly from each other. Our predictions are in  rough
agreement with the quark model calculations of Ref.~\cite{oh}, with
HQET+VMD results of Ref.~\cite{cfn} and with some of the predictions
of the QCD sum rules.~\footnote{The QCD sum rule results \cite{dn} for the
  $D^*$ decays were obtained using the $1/m_c$ expansion, which could
  be inaccurate due to the large value of the $1/m_c$ corrections.} It
is important to
note that in our calculations we do not need to introduce the anomalous
electromagnetic moment of the light quark as it is done in
Ref.~\cite{gr}, where it was found that its value should be rather
large ($\sim 0.5$) in order to get agreement with the experimental
(CLEO) value for $D^{*+}\to D^+\gamma$ decay rate.~\footnote{In
  Table~\ref{tab:2} we show predictions of Ref.~\cite{gr} for the
  value of anomalous electromagnetic quark moment equal to zero.}
The large value of the
anomalous electromagnetic moment is not justified phenomenologically (see
e.~g. Refs.~\cite{hs,gf}). The other differences of our calculations
from those of Ref.~\cite{gr} are the Lorentz structure of the confining
potential and a more comprehensive account of
relativistic effects. In particular, the relativistic transformation
of the meson wave function from the rest frame to a moving one given
by Eq.~(\ref{wig}) is missing in Ref.~\cite{gr}.          

\begin{sidewaystable}
 \caption{Comparison of different theoretical predictions for
   radiative decays of heavy-light mesons (in keV).} 
  \label{tab:2}
\centering
  \begin{tabular}{ccccccccc}
\hline
&\multicolumn{4}{c}{Quark models}&HQET+VMD&\multicolumn{3}{c}{QCD sum
  rules}\\
Decay& our&\cite{iv}&\cite{oh}&\cite{gr}&
\cite{cfn} &\cite{dn} &\cite{adip}& \cite{zyh}\\
\hline
$D^{* \pm}\to D^{\pm}\gamma$&1.04&0.36 &1.72 &0.050
&$0.51\pm0.18$ & $0.09^{+0.40}_{-0.07}$ &1.5& $0.23\pm0.1$\\
$D^{* 0}\to D^{0}\gamma$&11.5& 17.9& 7.18& 7.3&$16.0\pm7.5$
&$3.7\pm1.2$& 14.4& $12.9\pm2$\\
$D_s^*\to D_s\gamma$& 0.19&0.118& &0.101
&$0.24\pm0.24$ & &  &$0.13\pm0.05$\\
$B^{* \pm}\to B^{\pm}\gamma$&0.19&0.261& 0.272&0.084&
$0.22\pm0.09$ &$0.10\pm0.03$& 0.63& $0.38\pm0.06$\\
$B^{* 0}\to B^0\gamma$& 0.070&0.092 &0.064&0.037&
$0.075\pm0.027$ & $0.04\pm0.02$& 0.16&$0.13\pm0.03$\\
$B_s^*\to B_s\gamma$&0.054 & &0.051&0.035& & & &$0.22\pm0.04$\\
\hline
  \end{tabular}
\end{sidewaystable}

In Table~\ref{tab:3} we present the comparison of our results with the
predictions of different quark models \cite{eq,gklt,ful} for the 
rates of the radiative M1-transitions ($1^3S_1\to
1^1S_0+\gamma$) in the heavy-heavy $B_c$ meson. There we also give the
predicted values of the photon energy, which is determined by the mass
splitting of the vector and pseudoscalar ground states. In previous
calculations  \cite{eq,gklt,ful} the
nonrelativistic expression for the matrix element of the magnetic
moment was used. We see that even in the heavy-heavy $B_c$ meson
inclusion of the relativistic effects results in considerable
reduction of the radiative M1-decay rate. As can be seen from
Table~\ref{tab:1}, this reduction is evoked by significant
contributions of  relativistic effects for the $c$ quark, since
it is not heavy enough, as well as by the special choice of the mixture of
vector and scalar confining potentials in our model (\ref{vconf}).   

\begin{table}
\centering
 \caption{Comparison of theoretical predictions for the radiative
   $B^*_c\to B_c\gamma$ decay.} 
   \label{tab:3}
   \begin{tabular}{ccccc}
\hline
 & our& \cite{eq} &\cite{gklt} &\cite{ful}\\
\hline 
Photon Energy (MeV) & 61 & 72 & 64 & 55 \\
$\Gamma(B^*_c\to B_c\gamma)$ (eV) & 33 & 135 & 60 & 59 \\
\hline 
   \end{tabular}
   \end{table}

In summary we calculated radiative M1-decay rates of mesons with
open flavour in the framework of the relativistic quark model. In our
analysis the light quark was treated relativistically, while for the
heavy quark the $1/m_Q$ expansion was carried out up
to the second order. Relativistic consideration of
the light quark, relativistic heavy quark corrections as well as
Lorentz-structure of the confining potential  
considerably influence the predictions. We
find that only the mixture of vector and scalar
confining potentials (\ref{vconf}), with the mixing coefficient fixed
previously from quarkonium radiative decays \cite{gf} and weak decays
of heavy-light mesons \cite{fg}, is in agreement with recent CLEO data
for the $D^{*+}\to D^+\gamma$ decay rate. More precise measurement of
this decay rate and the measurement of radiative M1-decays of other
heavy-light mesons will be crucial for testing the relativistic quark
dynamics.    

The authors express their gratitude to M. M\"uller-Preussker and V. Savrin  
for support and discussions. We are grateful to S. Narison for the
useful comment on the QCD sum rule results.
Two of us (R.N.F and V.O.G.)
were supported in part by the {\it Deutsche
Forschungsgemeinschaft} under contract Eb 139/2-1, {\it
 Russian Foundation for Fundamental Research} under Grant No.\
00-02-17768 and {\it Russian Ministry of Education} under Grant
No. E00-3.3-45.

\end{document}